# Quantifying quantum coherence of multiple-charge states in tunable Josephson junctions

Jiangbo He[1,#], Dong Pan[2,#], Mingli Liu[1,3], Zhaozheng Lyu[1,4], Zhongmou Jia[1,3], Guang Yang[1], Shang Zhu[1,3], Guangtong Liu[1,4,5], Jie Shen[1,5], Sergey N. Shevchenko[6], Franco Nori[7,8,9,*], Jianhua Zhao[2,*], Li Lu[1,3,4,5,*], Fanming Qu[1,3,4,5,*]

[1] *Beijing National Laboratory for Condensed Matter Physics, Institute of Physics, Chinese Academy of Sciences, Beijing 100190, China*

[2] *State Key Laboratory of Superlattices and Microstructures, Institute of Semiconductors, Chinese Academy of Sciences, P.O. Box 912, Beijing 100083, China*

[3] *School of Physical Sciences, University of Chinese Academy of Sciences, Beijing 100049, China*

[4] *Hefei National Laboratory, Hefei 230088, China*

[5] *Songshan Lake Materials Laboratory, Dongguan 523808, China*

[6] *B. Verkin Institute for Low Temperature Physics and Engineering, Kharkiv 61103, Ukraine*

[7] *Theoretical Quantum Physics Laboratory, Cluster for Pioneering Research, RIKEN, Wakoshi, Saitama, 351-0198, Japan*

[8] *Quantum Computing Center, RIKEN, Wakoshi, Saitama, 351-0198, Japan*

[9] *Physics Department, The University of Michigan, Ann Arbor, MI 48109-1040, USA*

[#] These authors contributed equally to this work.

[*] Email: fnori@riken.jp; jhzhao@semi.ac.cn; lilu@iphy.ac.cn; fanmingqu@iphy.ac.cn

**Coherence and tunneling play central roles in quantum phenomena. In a tunneling event, the time that a particle spends inside the barrier has been fiercely debated. This problem becomes more complex when tunneling repeatedly occurs back and forth, and when involving many particles. Here we report the measurement of the coherence time of various charge states tunneling in a nanowire-based tunable Josephson junction; including single charges, multiple charges, and**

**Cooper pairs. We studied all the charge tunneling processes using Landau-Zener-Stückelberg-Majorana (LZSM) interferometry, and observed high-quality interference patterns. In particular, the coherence time of the charge states was extracted from the interference fringes in Fourier space. In addition, our measurements show the break-up of Cooper pairs, from a macroscopic quantum coherent state to individual particle states. Besides the fundamental research interest, our results also establish LZSM interferometry as a powerful technique to explore the coherence time of charges in hybrid devices.**

## Introduction

In a quantum tunneling event, a fundamental open question is the time spent by the particle(s) inside the barrier[1-6]. Another unsolved issue is the coherence time[7] of the back-and-forth tunneling process through the barrier, especially for multi-particles. Josephson junctions (JJs) provide a platform for studying tunneling processes of various charges; including single charges, coherent multiple charges, and Cooper pairs[8-10]. When a voltage is applied to a JJ, the well-known multiple Andreev reflections (MARs) provide a mechanism for the coherent transfer of multiple charges, up to infinity[11]. At the gap edge, the tunneling of single charges (Giaever tunneling) contributes to a coherence peak[12]. In addition, a JJ usually inherits the macroscopic quantum coherent nature of the superconductor. However, for an ultra-small JJ with a low capacitance and a weak Josephson coupling, the phase fluctuates significantly and is no longer a good quantum number, leading to a loss of macroscopic quantum coherence of Cooper pairs, regressing to individual particle states[9,12,13]. Accordingly, the Shapiro-step picture breaks down under microwave driving[14-16]. Therefore, a tunable JJ can serve as a unique platform for studying the coherence of tunneling processes of many charges, and for monitoring the disappearance and establishment of macroscopic quantum coherence of the Cooper pairs.

In recent years, scanning-tunneling-microscope (STM) experiments utilizing a superconducting tip have been used to distinguish the tunneling processes of the charges under microwave drive. The results were interpreted based on the Tien-Gordon theory and/or the microwave-assisted MAR model[13,17-20]. But the coherence time of various charges that tunnel back and forth through the barrier has not yet been examined. A systematic investigation of the onset and disappearance of macroscopic quantum coherence of the Cooper pairs in an ultra-small JJ is missing.

Interferometry can probe the coherence times of quantum states. Recently, Landau-Zener-Stückelberg-Majorana (LZSM) interferometry[21-23] has gained great success as a tool to diagnose the physical parameters and also to realize fast coherent manipulation of isolated qubits[24-31]. In this work, we implemented LZSM interferometry in a gate-tunable nanowire-based JJ, which is an open system (i.e., directly connected to the measurement circuit), in contrast to the generally isolated qubits. We constructed the JJ down to the single ballistic-channel limit, allowing us to resolve LZSM interference for single charges, multiple charges, and Cooper pairs —— in the weak coupling regime and under the drive of microwaves. Through analyzing the interference fringes in the two-dimensional (2D) Fourier space, for the first time we extracted the coherence times of all these charge states that tunnel back and forth through the junction. We further uncovered the loss of macroscopic quantum coherence of the Cooper pairs in the JJs.

**Results**

**LZSM interferometry.** Before presenting the experimental results, we first interpret the physical picture of LZSM interference in an ultra-small JJ. Phase fluctuations lead to the loss of macroscopic quantum coherence and Cooper pair tunneling as individual microscopic particles[9,12,13] (Fig.

1a). MARs enable a coherent transfer of 2-charges through a 1st-order reflection (Fig. 1b) and 3-charges through a 2nd-order reflection (Fig. 1c). Single charge tunneling is activated when the voltage reaches $V_b = 2\Delta/e$ ($\Delta$ is the superconducting gap and $e$ is the elementary charge) (Fig. 1d). However, to enable LZSM interferometry, two coupled energy levels plus a fast ac drive are required[32]. For a JJ, an open system, we argue that thanks to the singularities of the BCS single particle density of states and the Cooper pair condensate, the effective two levels can be mapped to the charges being at the left and the right sides of the junction, respectively. These two states, |L> and |R>, are shown in the bottom row of Figs. 1a-1d. The charges can tunnel between these two states under a microwave drive.

Figure 1e illustrates LZSM interference. The two states |L> (black dashed line) and |R> (green dashed line) anti-cross, with a minimal distance $\alpha$, which is twice the coupling strength. The junction is detuned by $V_0$ relative to the anti-crossing point, and a harmonic microwave (brown curve) with an amplitude of $V_{RF}$ drives the system adiabatically to and nonadiabatically through the anti-crossing at time $t = t_1$, with a transition probability of $P_{LZSM} = e^{(-\pi\alpha^2/2v\hbar)}$, where $v$ is the sweeping velocity. The two trajectories accumulate a phase difference (the shaded area) until $t = t_2$ when the system is brought back to the anti-crossing, where LZSM interference occurs.

The continuous microwave drive sustains the interference of the alternating tunneling until the charges lose coherence. To incorporate the environmental decoherence, classical noise can be introduced to the microwave drive, and a white noise model is employed under perturbation, to obtain the rate of transitions between |L> and |R> [33,34]

$$W(\varepsilon, A) = \frac{\alpha^2}{2} \sum_{n=-\infty}^{+\infty} \frac{\Gamma_2 \, J_n^2\left(\frac{A}{\hbar\omega}\right)}{(\varepsilon - n\hbar\omega)^2 + \hbar^2\Gamma_2^2}, \quad (1)$$

where $\varepsilon = meV_0$ is the detuning energy, $A = meV_{RF}$, $\omega = 2\pi f$ (with $f$ being the frequency of the microwave), $\Gamma_2 = 1/T_2$ is the decoherence rate, $n = \pm 1, \pm 2 \ldots$ denotes the satellite replicas of the $n = 0$ Lorentzian-shaped peaks, $J_n$ is the Bessel function of the first kind, and $m = 1, 2, 3 \ldots$ represents the number of charges in an elementary tunneling event. Evidently, the observability (broadening) of the interference fringes depends on the competition between ω and $\Gamma_2$.

The coherence time $T_2 = 1/\Gamma_2$ can be conveniently characterized in Fourier space by inverting the energy variable to the time variable. A 2D Fourier transform (2D FT) of $W(\varepsilon, A)$ yields[33]

$$W_{FT}(k_A, k_\varepsilon) = \frac{\alpha^2 \omega e^{-\Gamma_2 |k_\varepsilon|}}{2\sqrt{\frac{4}{\omega^2}\sin^2\left(\frac{1}{2}\omega k_\varepsilon\right) - k_A^2}} \quad (2)$$

for $|k_A| < \frac{2}{\omega}\left|\sin\left(\frac{1}{2}\omega k_\varepsilon\right)\right|$ and zero otherwise. The reciprocal-space variables $k_A$ and $k_\varepsilon$ correspond to the energy variables $A$ and $\varepsilon$, respectively. Therefore, lemon-shaped ovals following the singular boundary $\frac{\omega}{2}k_A = \pm\sin\left(\frac{1}{2}\omega k_\varepsilon\right)$ and an exponential decay on $k_\varepsilon$ as $e^{-\Gamma_2|k_\varepsilon|}$ are expected (Supplementary Information). Figure 1f illustrates a linear dependence of $\ln(W_{FT})$ on $|k_\varepsilon|$, whose slope generates the coherence time $T_2 = 1/\Gamma_2$.

**Epitaxial nanowire JJs.** Next, we present the experimental results. $InAs_{0.92}Sb_{0.08}$ nanowires were grown by molecular beam epitaxy, followed by an *in situ* epitaxy of ~15 nm-thick Al at a low temperature, which guarantees a hard superconducting gap[35,36]. Narrow Al gaps (JJs) were formed naturally during growth due to shadowing between the dense nanowires, eliminating the wet etching process and possible degradations. Figure 1g shows the nanowire possessing two JJs studied in this work, with Al gaps of ~20 nm (JJ1) and ~40 nm (JJ2), respectively. The yellow rectangles show post-fabricated Ti/Au (10/80 nm) contacts. Standard lock-in techniques and a microwave antenna were applied to carry out the transport measurements at ~10 mK. We will focus on JJ1 in the main text and present typical results for JJ2 in the Supplementary Information.

One of the advantages of the nanowire-based JJ is its full gate tunability. By applying a gate voltage $V_G$ (Figs. 2a and 2b), the JJ could be tuned to the strong coupling regime, the single ballistic-channel limit, and the weak-coupling regime. Figure 2a displays the correlation between the conductance $dI/dV$ measured at the normal state and the critical supercurrent $I_C$. A quantized conductance plateau at $2e^2/h$ corresponds to a single ballistic channel, and the associated $I_C$ stays around 15 nA, which is smaller than the theoretical expectation[37] of $e\Delta/\hbar = 52$ nA ($\Delta \approx 215$ μeV). The discrepancy could be explained by intrinsic reasons though[38], to the best of our knowledge, the value $I_C/\Delta$ in our device is the largest for a single channel[39-42]. Unlike the STM or the break-junction experiments where usually a mix of several channels contributes to the conductance[19,43], we demonstrate the single channel limit unambiguously, beneficial for the investigation of the coherence. Figure 2b displays the conductance map at the weak-coupling regime, showing the peaks contributed by Cooper pairs, 2-charges (the 1st-order MARs), and single charges, as marked by the green, red, and blue triangles, respectively.

**LZSM interference of single charges.** We proceed to turn on the microwave and examine the LZSM interferometry. To verify our implementation of the interferometry, the JJ was first set to the tunneling limit at $V_G = -31$ V. Since the conductance scales roughly to $m\tau^m$, where $\tau$ is the transmission[44], the conductance peaks of multiple charges were heavily suppressed for small $\tau$, leaving the single-charge coherence peaks as the prominent characteristics (Fig. 2b). Figure 2c depicts the $dI/dV$ versus microwave power $P$ and $V_b$ at a frequency $f = 11.755$ GHz, exhibiting clear interference fringes consistent with Eq. (1). After considering an attenuation of the microwave and converting $P$ to $V_{RF}$, the corresponding 2D FT is plotted in Fig. 2d, and the lemon-shaped ovals nicely follow the predictions of Eq. (2).

Figure 2e shows $\ln(W_{FT})$ as a function of $k_\varepsilon$, averaged near $k_A = 0$ to smooth out the singularities. Since $W_{FT} \propto e^{-\Gamma_2|k_\varepsilon|}$, a linear fit (red line) of the peak-values (blue squares) generates a coherence time $T_2 = 0.075$ ns. In addition, the Lorentzian shape of Eq. (1) versus $\varepsilon = meV_0$ depends on $T_2$ as well. Figure 2f illustrates a Lorentzian fit (red curve) of the $dI/dV$ versus $\Delta V_b = V_b - 2\Delta/e$ curve (black) at $P = -55$ dBm ($V_{RF} \approx 0$), which yields a similar $T_2 = 0.083$ ns.

Although $T_2$ is short, we demonstrate LZSM interferometry as a powerful approach to extract the coherence time of the charges in an open JJ. Two lower frequencies were further used, as shown in Figs. 2g-h for 7.665 GHz and Figs. 2i-j for 4.0 GHz. When the frequency decreases, the interference fringes become blurred since the time interval between two subsequent LZSM transitions increases and thus the interference loses coherence gradually[33]. The number of ovals in the Fourier space reveals a fingerprint of how many LZSM transitions (up to a factor of 2) can develop before decohering completely. Therefore, the observable ovals reduce from Fig. 2d, to Fig. 2h, and to Fig. 2j. Please refer to the Supplementary Information for a detailed data analysis.

**LZSM interference of various charges.** The JJ was then configured to a slightly stronger coupled regime ($V_G = -28.9$ V) to release the tunneling of multiple charges and Cooper pairs. Figure 3a presents the interference fringes of the Cooper pairs, 2-charges through the 1$^{st}$-order MARs and single charges, as marked by the square, circle, and triangle, respectively. The fringes of 3-charges through the 2$^{nd}$-order MARs can also be recognized by the dashed lines in Fig. 3b, a zoom-in of Fig. 3a. The 2D FT maps were plotted in Figs. 3d-f, marked in correspondence with Fig. 3a, and $T_2$ were extracted to be 0.1 ns, 0.049 ns, and 0.051 ns, for Cooper pairs, 2-charges, and single charges, respectively.

Taking the d$I$/d$V$ versus $V_b$ line-cut at $P = -45$ dBm in Fig. 3a as an input and using the $T_2$ values, the conductance map can be calculated. For this we assume that d$I$/d$V$ is proportional to $W$ and its behavior follows Eq. (1). This is shown in Fig. 3c, which manifests a good agreement with the measured data in Fig. 3a. So far, we realized the measurement of the coherence times for the tunneling of all these charges.

We next discuss about three more characteristics of the interference fringes. First, the number, $m$, of charges can be conveniently determined from the spacing ($\delta V_b$) of the satellite peaks, since $\delta V_b \propto 1/m$. Figure 3g plots d$I$/d$V$ versus the $V_b$ curves taken from Figs. 3a and 3b, as indicated by the black and red bars. The spacings $\delta V_b$ of 49 μV, 25 μV, 15 μV and 24 μV are consistent with charge numbers $m$ of 1, 2, 3 and 2 at $V_b = -2\Delta/e$, $-2\Delta/(2e)$, $-2\Delta/(3e)$ and 0, respectively.

Second, the d$I$/d$V$ peaks behave as $J_n^2$, versus the microwave amplitude $V_{RF}$. In Fig. 3h, the left and right columns show the extracted curves (black) from Fig. 3a for the main peak ($n = 0$) and the first satellite peak ($n = 1$), respectively. A good agreement was achieved by a $J_n^2$ fitting, from Eq. (1), as shown by the red curves. Note that a d$I$/d$V$ constant was included in each fitting to account for the background conductance, and the power $P$ has been converted to $V_{RF}$. Moreover, the $J_n^2$ dependence serves as evidence for Cooper pair tunneling as microscopic particles near $V_b = 0$, instead of the macroscopic Shapiro-step picture whose characteristic width would rather follow $J_n$ to the 1st-order[13,19], as shown below.

**Onset and loss of the macroscopic coherence.** Third, there is an opposite trend of the coherence for the Cooper pairs and the other charges (quasiparticles) when the coupling strength of the JJ increases. A clue can be found by examining the $T_2$ extracted above. When increasing the coupling strength of the JJ from $V_G = -31$ V to $-28.9$ V, $T_2$ for single charges drops from 0.075 ns to 0.051 ns. At $V_G = -28.9$ V, $T_2 = 0.049$ ns and 0.1 ns for 2-charges through MARs, and Cooper pairs, respectively. When the coupling strength increases further, as shown in Fig. 3i at $V_G = -27.5$ V, the fringes for the Cooper pairs become more evident while the others fade away. This is consistent with the scenario that when the Josephson coupling and the related capacitance increase, the Cooper pairs tend to recover the macroscopic quantum coherence, but the quasiparticles decohere faster due to the energy broadening and the background conductance, etc. Eventually, at the strong-coupling regime the macroscopic quantum coherence was fully restored and regular Shapiro steps were obtained at $V_G = 6$ V, and a good agreement was achieved by fitting the step width to the 1st-order of the Bessel function, $|J_n|$, as shown in Fig. 4.

To conclude, we realized the measurement of the coherence time of various charge states that tunnel back and forth in a JJ by implementing LZSM interferometry. Then, we revealed the change of the Cooper pairs in the JJ, from a macroscopic quantum coherent state to individual particle states; or conversely, the onset of macroscopic quantum coherence. We expect that the LZSM interferometry is a powerful technique for exploring the coherence time of charges tunneling in various hybrid devices, such as through trivial Andreev and Yu-Shiba-Rusinov bound sates[20], Floquet states[45], as well as topological Majorana bound states[46-49]. Note that these states might also be subjected to the crossover of the Cooper pairs from a macroscopic quantum coherence state to individual particle states.

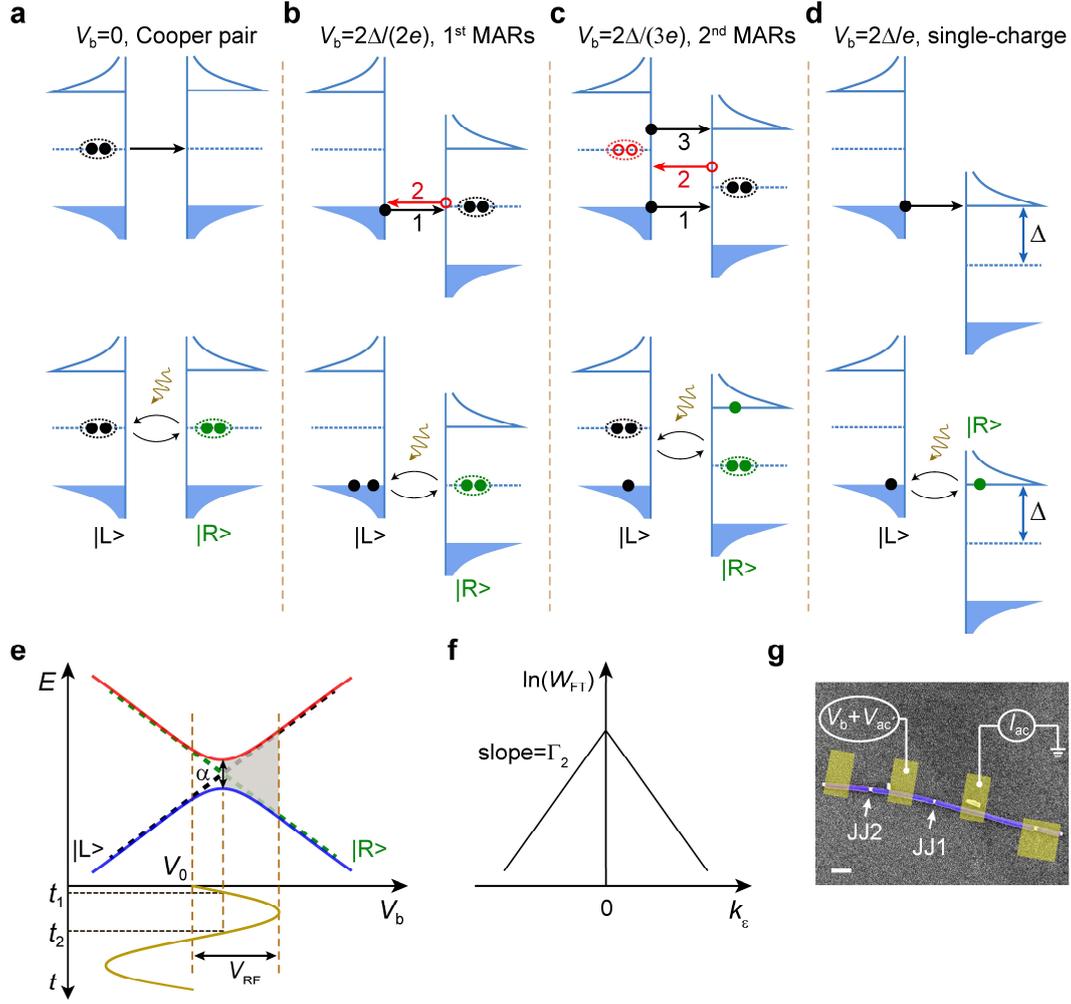

**Figure 1. Illustration of LZSM interference in a Josephson junction. a-d,** Charge tunneling at typical bias voltages $V_b$. The arrows in **b** and **c** indicate successive Andreev reflections. |L> and |R> represent the two effective states where charges tunnel back and forth under a microwave drive. **e,** The dashed lines show the diabatic energy levels for the states |L> and |R>; the solid blue and red curves show the energy levels when taking tunneling into account. A microwave (brown) drives the system through the anti-crossing at the time $t_1$ where LZSM transitions occur, and back to the anti-crossing at time $t_2$, where the two trajectories interfere. **f,** Linear dependence of $\ln(W_{FT})$ on $|k_\varepsilon|$, whose slope is $\pm\Gamma_2$. **g,** The nanowire (white) contains JJ1 and JJ2. Aluminum is shown in violet, and the yellow rectangles denote post-fabricated Ti/Au contacts. Scale bar: 200 nm.

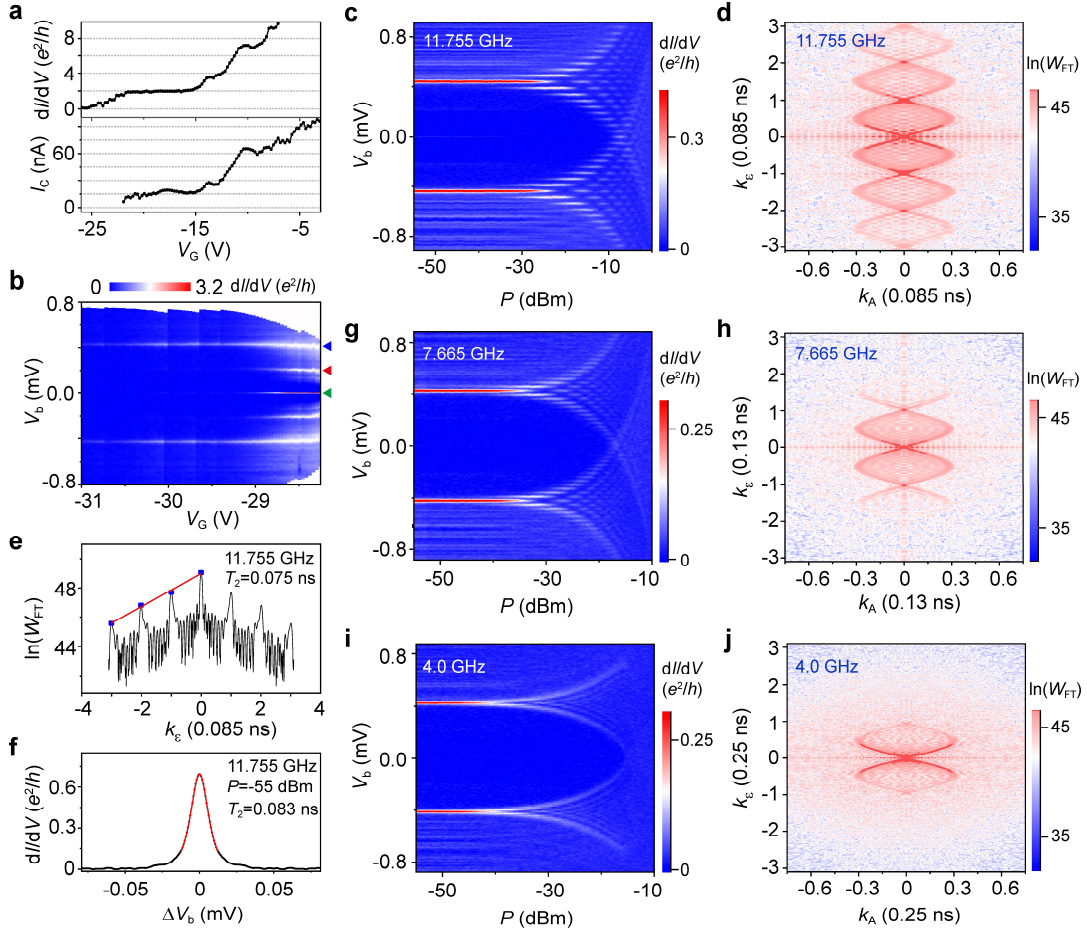

**Figure 2. LZSM interference in the tunneling limit. a**, Conductance d$I$/d$V$, measured in the normal state, and critical supercurrent $I_c$, both versus the gate voltage $V_G$. **b**, d$I$/d$V$ versus $V_G$ and $V_b$ at the weak coupling regime. The green, red and blue triangles mark the peaks of Cooper pairs, 2-charges of the 1$^{st}$-order MARs, and single charges, respectively. **c**, Interference fringes at $V_G = -31$ V and $f = 11.755$ GHz versus microwave power $P$ and $V_b$. **d**, The 2D FT of **c**. For clarity, $k_A$ and $k_\varepsilon$ are in units of 1/$f$. **e**, Averaged line-cut (black) near $k_A = 0$ of **d**. The red line is a linear fit of the peak-values as marked by the blue squares. **f**, Lorentzian fit (red curve) of the d$I$/d$V$ data (black curve) versus $\Delta V_b = V_b - 2\Delta/e$ at $P = -55$ dBm of **c**. The data for $\Delta V_b < 0$ are extracted from **c**, and the $\Delta V_b > 0$ branch is simply mirrored. **g-h (i-j)**, Interference fringes and the corresponding 2D FT at $f =$ 7.665 GHz (4.0 GHz).

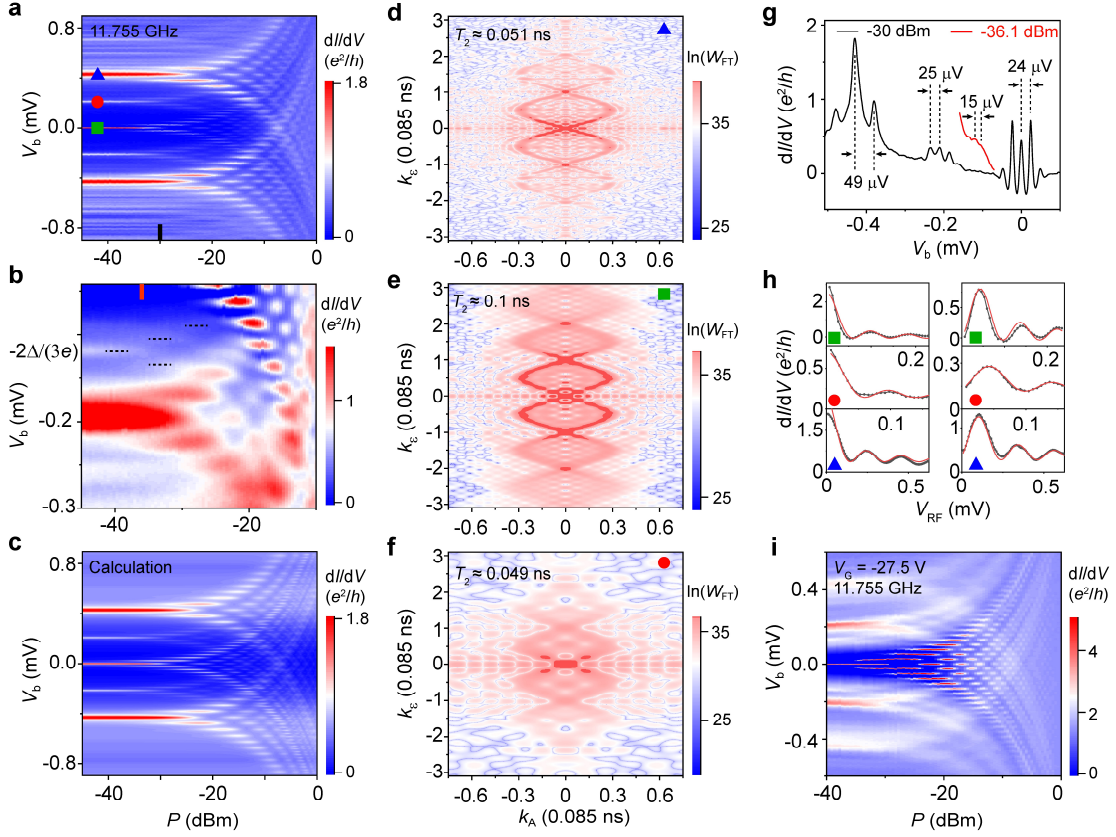

**Figure 3. LZSM interference of various charges. a**, Interference fringes at $V_G = -28.9$ V and $f = 11.755$ GHz. The square, circle and triangle depict the contributions of Cooper pairs, 2-charges through the 1st-order MARs, and single charges, respectively. **b**, Zoom-in of **a**. The dashed lines highlight the features of 3-charges through the 2nd-order MARs. **c**, Calculated $dI/dV$ based on both the $dI/dV$ versus $V_b$ curve at $P = -45$ dBm in **a** and the $T_2$ values shown in **d-f**, which display the 2D FT maps of the sets of the fringes in **a**. **g**, Line-cuts taken from **a** and **b**, as indicated by the black and red bars. **h**, Extracted curves (black) from the fringes of **a** and the $J_n^2$ fit (red). The left column is for the $n = 0$ main peaks, and the right for the $n = 1$ satellite peaks. **i**, Interference fringes at $V_G = -27.5$ V and $f = 11.755$ GHz.

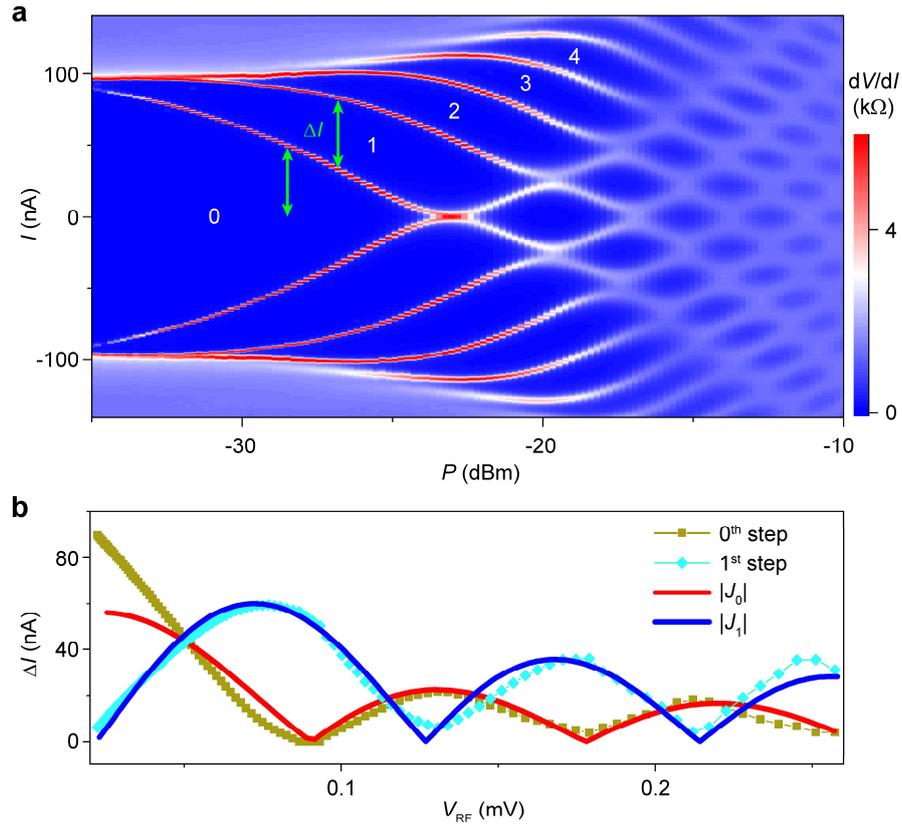

**Figure 4. Shapiro steps in the strong-coupling regime. a**, Differential resistance d$V$/d$I$ versus $P$ and current $I$, measured at $V_G$ = 6 V and $f$ = 11.56 GHz. The numbering indicates the Shapiro steps. **b**, Step width $\Delta I$ versus $V_{RF}$ and the corresponding Bessel function fitting to the 1$^{st}$-order. A −45 dB attenuation was assumed and subtracted.

## Methods

*InAs$_{0.92}$Sb$_{0.08}$-Al Nanowire Growth*: InAs$_{0.92}$Sb$_{0.08}$ nanowires were grown in a solid source molecular beam epitaxy system (VG V80H) on p-Si (111) substrates using Ag as catalysts[50]. The nanowires were grown for 40 min at a temperature of 465 °C with the beam fluxes of In, As and Sb sources of 1.1×10$^{-7}$ mbar, 4.6×10$^{-6}$ mbar and 4.7×10$^{-7}$ mbar, respectively. After the growth of InAs$_{0.92}$Sb$_{0.08}$ nanowires, the sample was transferred from the growth chamber to the preparation chamber at 300 °C to avoid arsenic condensation on the nanowire surface. The sample was then cooled down to a low temperature (~ –30 °C) by natural cooling and liquid nitrogen cooling[51]. Al was evaporated from a Knudsen cell at an angle of ~20° from the substrate normal (~70° from the substrate surface) and at a temperature of ~1150 °C for 180 s (giving approximately 0.08 nm/s). During the Al growth, the substrate rotation was kept disabled. When the growth of nanowires with Al was completed, the sample was rapidly pulled out of the MBE growth chamber and oxidized naturally.

*Fabrication of the devices*: The as-grown nanowires were transferred onto Si/SiO$_2$ (300 nm) substrates simply by a tissue. Standard electron-beam lithography was applied to fabricate the Ti/Au (10 nm/80 nm) electrodes using electron-beam evaporation. Note that an ion source cleaning and a soft plasma cleaning were performed to improve the contact prior to the deposition.

*Transport measurements*: The measurements were carried out in a cryo-free dilution refrigerator at a base temperature of ~10 mK. Low-frequency lock-in techniques were utilized to measure the conductance of the JJ, $dI/dV \equiv I_{ac}/V_{ac}$, with a small excitation ac voltage $V_{ac}$ and a dc bias voltage $V_b$ applied and the ac current $I_{ac}$ measured. The microwave driving was supplied through a semi-rigid coaxial cable with an open end which is several millimeters above the JJ, serving as an antenna.

*Data analysis*: The measured LZSM interference for the various charges in our small JJ manifests as sets of d$I$/d$V$ fringes as a function of microwave power $P$ and bias voltage $V_b$. However, $P$ is the output of the signal generator, and the effective power $P_{eff}$ on the JJ needs to be determined. To do so, we calculated the interference fringes for a given measured data set using Eq. (1), and compare their power difference of the $n = 0$ peaks to extract the attenuation of the microwave, i.e., $P$–$P_{eff}$. Afterwards, the measured interference fringes were selected and converted from power ($P_{eff}$) dependence to amplitude ($V_{RF}$) dependence. And the data set was further symmetrized to the four quadrants and scaled to a maximum of 1 to carry out the 2D FT. A detailed interpretation can be found in the Supplementary Information.

**Acknowledgements**

We would like to thank Q. F. Sun, Y. Zhou, B. Lu, and N. Kang for fruitful discussions, and also thank L. J. Wen, L. Liu, R. Zhuo and F. Y. He for assistance in sample growth. This work was supported by the National Key Research and Development Program of China through Grants Nos. 2022YFA1403400 and 2017YFA0304700; by the NSF China through Grants Nos. 12074417, 92065203, 92065106, 61974138, 11874406, 11774405, and 11527806; by the Strategic Priority Research Program of Chinese Academy of Sciences, Grants Nos. XDB28000000 and XDB33000000; by the Synergetic Extreme Condition User Facility sponsored by the National Development and Reform Commission; by the Innovation Program for Quantum Science and Technology, Grant No. 2021ZD0302600. D.P. also acknowledges the support from Youth Innovation Promotion Association, Chinese Academy of Sciences (Nos. 2017156 and Y2021043). S.N.S. also acknowledges the support from the Army Research Office under Grant No. W911NF-20-1-0261. F.N. also acknowledges the support from Nippon Telegraph and Telephone

Corporation (NTT) Research, the Japan Science and Technology Agency (JST) [via the Quantum Leap Flagship Program (Q-LEAP), and the Moonshot R&D Grant Number JPMJMS2061], the Japan Society for the Promotion of Science (JSPS) [via the Grants-in-Aid for Scientific Research (KAKENHI) Grant No. JP20H00134], the Asian Office of Aerospace Research and Development (AOARD) (via Grant No. FA2386-20-1-4069), and the Foundational Questions Institute Fund (FQXi) via Grant No. FQXi-IAF19-06.**Author contributions**

F.Q. conceived and supervised the project. D.P. grew and characterized the nanowires. J.Z. supervised the nanowire growth. J.H. and M.L. fabricated the devices and performed the transport measurements. Z.L., Z.J., G.Y., S.Z., G.L., J.S., L.L. and F.Q. supported the fabrication and measurements. J.H. and F.Q. analyzed the data with the help from Z.L., G.Y., S.N.S. and F.N. F.Q., J.H., D.P., L.L., S.N.S. and F.N. wrote the manuscript, with input from all authors.

**Competing interests**

The authors declare no competing interests.

**Additional information**

Supplementary Information: The online version contains supplementary material available at [URL].

**Data availability**

The data supporting the findings of this study are available from the corresponding authors upon reasonable request.

Supplementary Information for

# Quantifying quantum coherence of multiple-charge states in tunable Josephson junctions

**Contents**

**Section 1. Growth of the epitaxial nanowires.**

In this work, all the nanowires were grown on p-type Si (111) substrates. The <111> and non-<111>-oriented InAs$_{0.92}$Sb$_{0.08}$-Al nanowires coexisted on the substrate surface because the thin natural oxide layer on the Si substrates cannot be removed completely before the nanowire growth. According to detailed transmission electron microscope (TEM) observations, we find that continuous half Al shells can be successfully grown on the facets of InAs$_{0.92}$Sb$_{0.08}$ nanowires with non-<111> growth directions, while the Al shells are discontinuous and look like 'pearls on a string' on the side walls of the <111> oriented nanowires. Particularly, narrow Al gaps can form naturally in these non-<111> InAs$_{0.92}$Sb$_{0.08}$-Al nanowires due to shadowing between the dense nanowires. Figure S1 shows a high-angle annular dark-field scanning transmission electron microscope (HAADF-STEM), energy dispersive spectrum (EDS) and high-resolution TEM data of a typical shadow InAs$_{0.92}$Sb$_{0.08}$-Al nanowire. As shown in Fig. S1a, a half Al shell with a narrow Al gap can be clearly observed on the facet of the InAs$_{0.92}$Sb$_{0.08}$ nanowire. The false-color EDS elemental maps of In (Fig. S1b), As (Fig. S1c), Sb (Fig. S1d) and Al (Fig. S1e) of the InAs$_{0.92}$Sb$_{0.08}$-Al nanowire further confirm that a narrow Al gap indeed formed in the continuous Al shell. Figures S1f and S1g are high-resolution TEM images of the nanowire taken from the InAs$_{0.92}$Sb$_{0.08}$ region and InAs$_{0.92}$Sb$_{0.08}$-Al interface area, respectively. The InAs$_{0.92}$Sb$_{0.08}$ nanowire has a pure zinc-blende crystal structure due to its non-<111> growth direction, although the Sb content is low, which is consistent with our previous work[1]. As shown in Fig. S1g, an abrupt interface between the Al shell and the InAs$_{0.92}$Sb$_{0.08}$ nanowire can be observed.

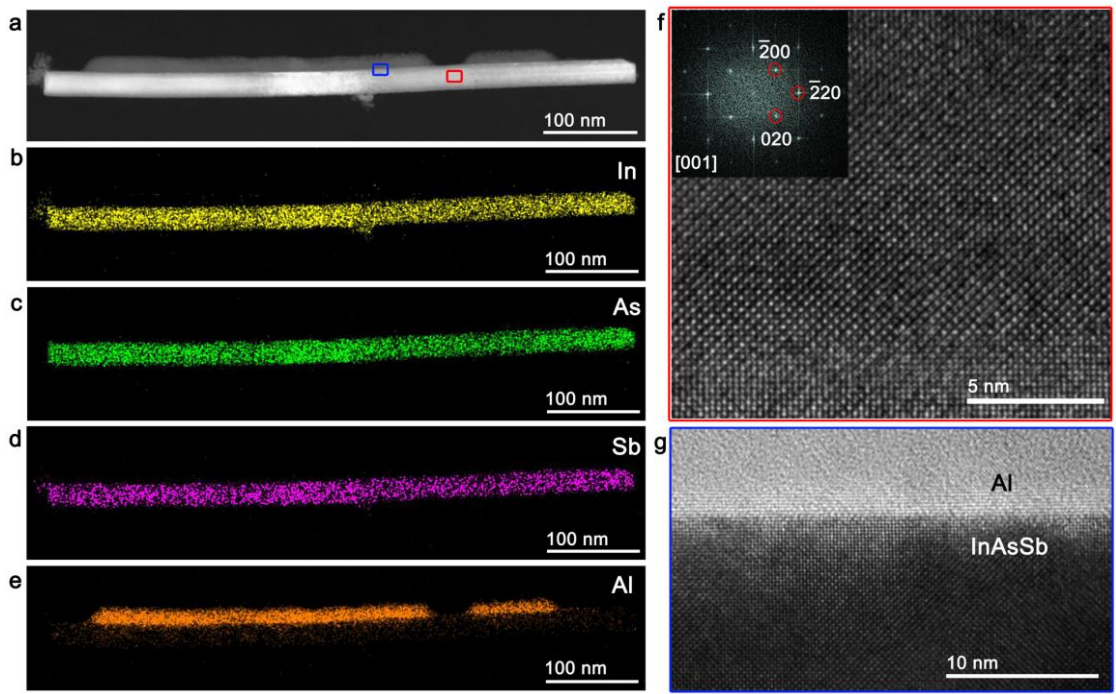

**Figure S1. Chemical composition and crystal structure of the InAs$_{0.92}$Sb$_{0.08}$-Al nanowires. a,** HAADF-STEM image of an InAs$_{0.92}$Sb$_{0.08}$-Al nanowire. **b-e,** False-color EDS elemental maps of In (yellow), As (green), Sb (purple), and Al (orange) of the InAs$_{0.92}$Sb$_{0.08}$-Al nanowire, respectively. **f,** High-resolution TEM image of the InAs$_{0.92}$Sb$_{0.08}$ nanowire. The inset of **f** is its corresponding fast Fourier transform. **g,** High-resolution TEM image of the InAs$_{0.92}$Sb$_{0.08}$ nanowire taken from the InAs$_{0.92}$Sb$_{0.08}$-Al interface area. The blue and red rectangles in **a** highlight the regions where high-resolution TEM images were recorded.

## Section 2. Measurement setup.

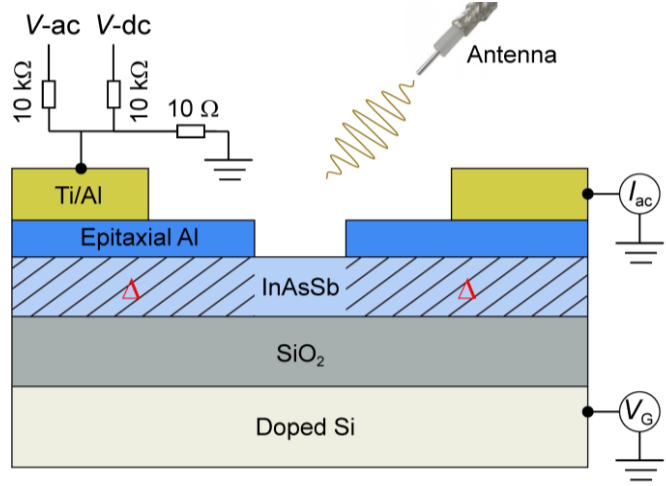

**Figure S2. The measurement setup.**

Figure S2 shows the transport measurement setup. A small ac voltage $V$-ac and a dc voltage $V$-dc were applied through a voltage divider, after which the actual $V_{ac}$ and $V_b$ were subjected onto the device. The ac current $I_{ac}$ was measured. A serial resistance from the dc wires and the low-pass filters was subtracted during the data analysis. A gate voltage $V_G$ was applied through a 300 nm-thick $SiO_2$ to control the coupling strength of the Josephson junction (JJ). The shaded InAsSb nanowire segments were brought into proximity with the epitaxial Al to possess a hard superconducting gap of size $\Delta$. A microwave antenna with an open end of a coaxial cable was implemented to radiate the junction with microwaves. For the critical current measurement and the Shapiro step measurement, the dc current was converted.

## Section 3. Theoretical treatment of the LZSM interference.

In this section, we interpret the theoretical treatment of the LZSM interference in the small JJ studied in our work. For ease of reading, we illustrate the LZSM interference again in Figs. S3a-S3c. As explained in the main text, we map the effective two states as the charges being on the left and the right side of the junction based on the Bardeen-Cooper-Schrieffer singularities of the density of states (see Fig. S3a for the case of Cooper pairs). The coupling strength between the two states, |L> and |R>, is defined as $\alpha/2$, resulting in an anti-crossing of $\alpha$. The red and blue energy levels in Fig. S3b are thus the two effective levels for LZSM interference. We consider a harmonic microwave driving with an amplitude of $V_{RF}$ and a frequency of $f$. The detuning can be controlled by applying a bias voltage relative to the anti-crossing point, $V_0$. As displayed in Figs. S3b and S3c, a LZSM transition takes place at time $t_1$ and (after accumulating a phase difference of $\theta$) a subsequent LZSM transition and LZSM interference occur at time $t_2$, when the system is driven back to the anti-crossing point.

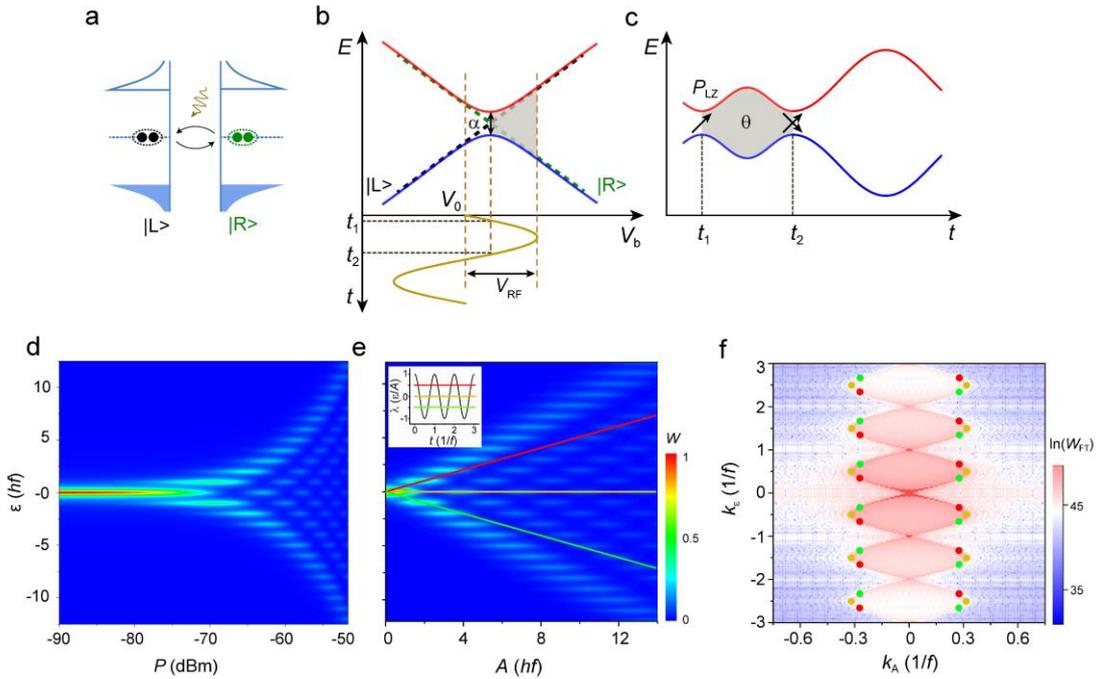

**Figure S3. LZSM interference and its Fourier transform. a**, Illustration of the two effective states in a small JJ. **b, c,** Schematics of the two-level system and the microwave driven LZSM interference. **d, e,** Calculated interference fringes. A harmonic driving signal (inset in **e**) with a frequency $f = 10$ GHz, and a decoherence rate $\Gamma_2 = 1/(4\omega)$ ($\omega = 2\pi f$) were used. **f**, The two-dimensional Fourier transform (2DFT) of **e**. The colored dots stand for the corresponding contributions from the lines in **e**.

The extraction of the coherence time in Fourier space for a driven qubit has been theoretically studied by Rudner et al[2]. Here we follow a similar procedure and theoretically analyze the LZSM interference for various charges in our small JJs. When a harmonic drive $V_{RF}\cos(\omega t)$ ($\omega = 2\pi f$) is applied, the Hamiltonian of the system can be expressed as:

$$H = -\frac{\hbar}{2}\begin{pmatrix} \beta(t) & \alpha \\ \alpha & -\beta(t) \end{pmatrix}, \quad \beta(t) = \varepsilon - A\cos(\omega t), \quad \text{(S1)}$$

where $\varepsilon = meV_0$, $A = meV_{RF}$, and $m$ is the number of charges, and $e$ is the elementary charge. When $V_{RF} > V_0$, the system will be driven to pass through the anti-crossing twice at $t_1$ and $t_2$ in one period, respectively. The times $t_{1,2}$ satisfy

$$A\cos(\omega t_1) = A\cos(\omega t_2) = \varepsilon, \quad \left(-1 < \frac{\varepsilon}{A} < 1\right). \quad \text{(S2)}$$

The phase difference (gray shaded area in Fig. S3c) accumulated between the two LZ transitions can be expressed as:

$$\theta(A, \varepsilon) = \int_{t_1}^{t_2} \beta(t)dt = \varepsilon(t_2 - t_1) - A\int_{t_1}^{t_2} \cos(\omega t)dt. \quad \text{(S3)}$$

Let us define a wave vector in the time domain

$$(k_A, k_\varepsilon) = \pm\left(\nabla_A \theta(A, \varepsilon), \nabla_\varepsilon \theta(A, \varepsilon)\right), \quad \text{(S4)}$$

where $\pm$ accounts for the contributions of $e^{\pm i\theta(A,\varepsilon)}$. Considering that the net contributions of $\nabla_A t_{1,2}$ and $\nabla_\varepsilon t_{1,2}$ vanish due to Eq. (S2),

$$(k_A, k_\varepsilon) = \pm\left(-\int_{t_1}^{t_2} \cos(\omega t)\,dt, t_2 - t_1\right). \quad \text{(S5)}$$

Obviously, $k_A$ and $k_\varepsilon$ correspond to the phase gain and the time separation, respectively. The solution of Eq. (S2), $t_2 = -t_1 = \frac{1}{\omega}\arccos(\varepsilon/A)$, gives $\int_{t_1}^{t_2} \cos(\omega t)dt = (2/\omega)\sqrt{1 - \varepsilon^2/A^2}$. Therefore, Eq. (S5) defines a curve that is dependent only on the parameter $\lambda = \varepsilon/A$. Substituting these results into Eq. (S5), we obtain that the parametric curve

is a sine function

$$\omega \frac{k_A}{2} = \pm \sin\left(\frac{\omega k_\varepsilon}{2}\right). \tag{S6}$$

As explained later, this function corresponds to the lemon-shaped ovals with a singular boundary in the 2DFT $(k_A, k_\varepsilon)$ space of the interference fringes (see also Fig. S3f).

To account for decoherence, a classical noise can be added to the microwave drive, $\tilde{\beta}(t) = \beta(t) + \delta\varepsilon(t)$. The rate of transitions between $|L\rangle$ and $|R\rangle$ can be easily deduced in a rotating frame, where we have

$$H = -\frac{\hbar}{2}\begin{pmatrix} 0 & \alpha(t) \\ \alpha^*(t) & 0 \end{pmatrix}, \qquad \alpha(t) = \alpha e^{-i\tilde{\theta}(t)}, \tag{S7}$$

with $\tilde{\theta}(t) = \int_0^t \tilde{\beta}(t')dt'$. The rate of transitions between $|L\rangle$ and $|R\rangle$ is [2]

$$W = \lim_{\delta t \Gamma_2 \gg 1} \frac{\alpha^2}{4\delta t} \iint_t^{t+\delta t} \langle e^{-i\tilde{\theta}(t_1)} e^{i\tilde{\theta}(t_2)} \rangle_{\delta\varepsilon} \, dt_1 dt_2, \tag{S8}$$

where $\Gamma_2 = \frac{1}{T_2}$ is the decoherence rate. A white noise model can be applied to average over $\delta\varepsilon(t)$: $\langle e^{i\delta\theta(t_2) - i\delta\theta(t_1)}\rangle_{\delta\varepsilon} = e^{-\Gamma_2|t_1 - t_2|}$, where $\delta\theta(t) = \int_0^t \delta\varepsilon(t')dt'$.

The Fourier series expansion, $e^{i\theta(t)} = e^{i\varepsilon t}\sum_n J_n\left(\frac{A}{\omega}\right) e^{-in\omega t}$, where $J_n$ is a Bessel function of the first kind, can be used to obtain

$$W(\varepsilon, A) = \frac{\alpha^2}{2} \sum_{n=-\infty}^{+\infty} \frac{\Gamma_2 J_n^2\left(\frac{A}{\hbar\omega}\right)}{(\varepsilon - n\hbar\omega)^2 + \hbar^2\Gamma_2^2}. \tag{S9}$$

It is this equation, Eq. (S9), that relates to the conductance of the JJ, a measurable parameter in the experiment. It describes the interference fringes as a function of the microwave drive $A = meV_{RF}$ and the detuning (voltage bias) $\varepsilon = meV_0$ relative to the anti-crossing point. Figure S3e presents one example calculated directly from Eq. (S9), using a frequency $f = 10$ GHz and a decoherence rate $\Gamma_2 = 1/(4\omega)$. Note that the data has been scaled to a maximum of 1, and the x-axis is the microwave amplitude $V_{RF}$ ($A = meV_{RF}$), while in our measured data it is the power $P$. In addition, along the y-axis, the spacing $\delta V_b$ of the satellite peaks satisfies $hf = me\delta V_b$, which enables a direct extraction of the number of charges $m$ for the various tunneling processes in our experiment.

The Fourier transform, $W_{FT}(k_A, k_\varepsilon) = \int_{-\infty}^{+\infty} e^{-ik_A A - ik_\varepsilon \varepsilon} W(A, \varepsilon) \, dA \, d\varepsilon$, maps $W(\varepsilon, A)$ from the energy space to the Fourier space in the time domain[2],

$$W_{FT}(k_A, k_\varepsilon) = \frac{\alpha^2 \omega \, e^{-\Gamma_2 |k_\varepsilon|}}{2\sqrt{\frac{4}{\omega^2} \sin^2\left(\frac{1}{2}\omega k_\varepsilon\right) - k_A^2}}. \quad (S10)$$

We can see that $W_{FT}(k_A, k_\varepsilon)$ is concentrated inside the region bounded by the sinusoids $\frac{\omega}{2} k_A = \pm \sin\left(\frac{1}{2}\omega k_\varepsilon\right)$, which is the Eq. (S6) shown above. Therefore, lemon-shaped ovals following such singular boundary are expected in Fourier space. Crucially, an exponential decay on $k_\varepsilon$ as $e^{-\Gamma_2 |k_\varepsilon|}$ can help to directly extract the coherence time $T_2$, which demonstrates that this is a powerful technique.

Figure S3f shows the 2DFT results of Fig. S3e. Note that the integral to obtain $W_{FT}$ is from $-\infty$ to $+\infty$; therefore, the fringes shown in Fig. S3e need to be symmetrized to the four quadrants before performing the 2DFT. Lemon-shaped ovals are consistent with Eq. (S10) and also with our experimental results. In addition, for a given $\lambda$, there is a ray $\varepsilon = \lambda A$ in energy space $(A, \varepsilon)$ and a set of periodic points in the temporal space. As shown in Figs. S3e and S3f, the colored rays of the interference fringes contribute to the colored dots in the Fourier space correspondingly. For example, the red ray stands for $\varepsilon = \frac{1}{2}A$, and accordingly,

$$(k_A, k_\varepsilon) = \pm\left(-\left(\frac{2}{\omega}\right)\sqrt{1-\frac{\varepsilon^2}{A^2}}, \frac{2}{\omega}\arccos(\frac{\varepsilon}{A})\right) = \pm\left(-\frac{\sqrt{3}}{2\pi f}, \frac{1}{3f}\right).$$

**Section 4. Choice of the microwave frequencies.**

In order to choose proper frequencies of the microwave, we measured the d$I$/d$V$ dependence on both the microwave frequency $f$ and power $P$ at a back-gate voltage $V_G = -30$ V and a bias voltage near the gap edge, as shown in Fig. S4. There is a clear frequency-dependent effective attenuation of the microwave, presumably due to the coaxial lines and the details of the coupling between the microwave and the device. Nevertheless, in order to get rid of the heat load on the dilution refrigerator, we selected frequencies corresponding to a local minimum attenuation, as marked by the green arrows for 7.665 GHz and 11.755 GHz, respectively.

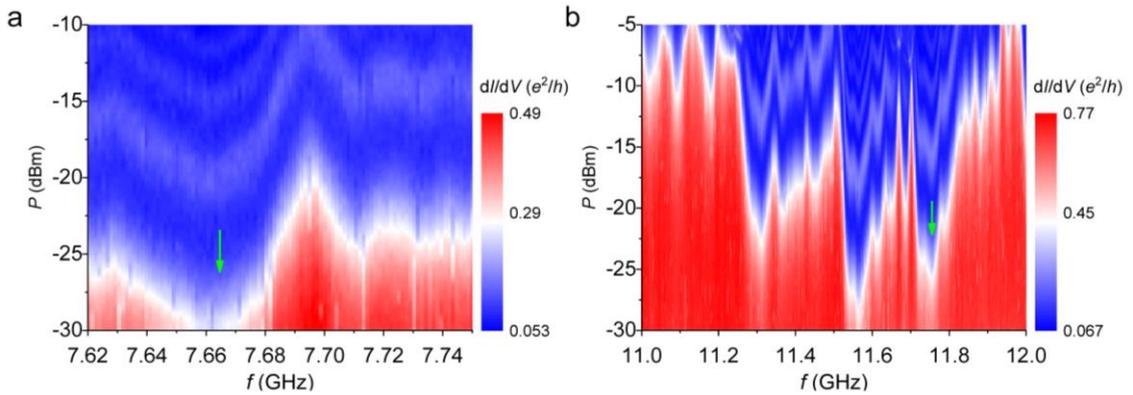

**Figure S4. Choice of the microwave frequencies.** The d$I$/d$V$ was measured as a function of the microwave frequency $f$ and power $P$ using a back-gate voltage $V_G = -30$ V and a bias voltage near the gap edge.

## Section 5. Data analysis

In this section, we explain the data analysis process of the interference fringes. As an example, here we study the data shown in Fig. 2c in the main text for $V_G = -31$ V and $f = 11.755$ GHz.

**Step 1**: determine the attenuation of the microwave.

As shown in Section 4, there is a frequency-dependent attenuation of the microwave. Therefore, we need to determine the actual microwave power experienced by the junction for each frequency used in the experiment. For instance, by applying Eq. (S9) we calculated the interference fringes for single charges at $V_G = -31$ V and $f = 11.755$ GHz, corresponding to Fig. 2c in the main text, as shown in Fig. S5a. The coherence time $T_2$ was assumed to be 0.1 ns, and the peak conductance was set to be the same as the measurement. The abscissa (horizontal axis) was converted between $P$ and $V_{RF}$ using the relation:

$$V_{RF} = \sqrt{\frac{50 * 10^{\frac{P}{10}}}{1000}}, \qquad (S11)$$

where $P$ is in dBm and $V_{RF}$ in volts, and 50 Ω is the impedance. As plotted in Fig. S5b, the attenuation at this frequency can be achieved by a comparison between the calculated curve (green) and the measured curve (black). The two arrows mark the positions of the first peaks, and the attenuation was $-61.6 + 16.1 = -45.5$ (dB).

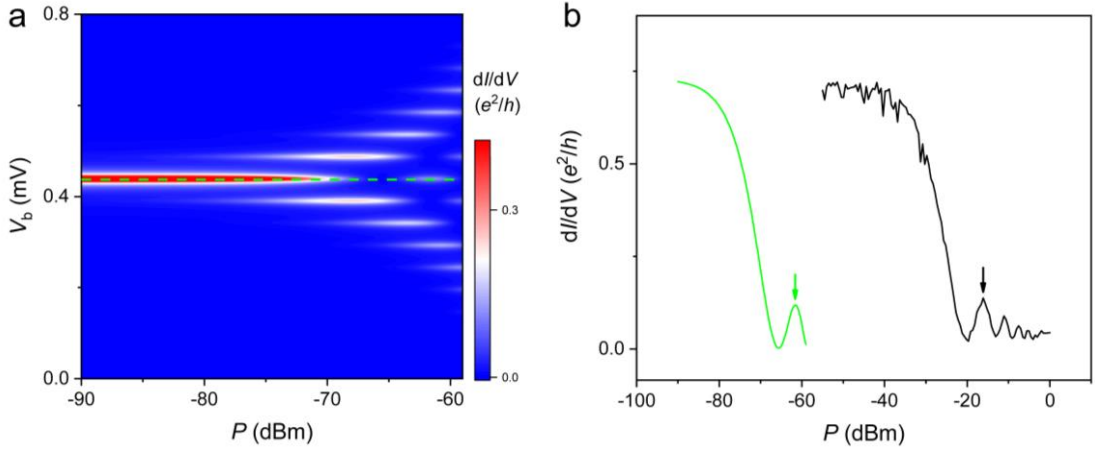

**Figure S5. Determination of the attenuation. a**, Calculated interference fringes corresponding to Fig. 2c in the main text, without an attenuation of the microwave. **b**, Comparison between the experimental data (black) taken from Fig. 2c in the main text and the calculated curve taken from **a**, as indicated by the green dashed line. An attenuation of −45.5 dB can be extracted by the power difference between the first peaks, as marked by the green and black arrows.

**Step 2**: select the proper data set and convert to the four quadrants.

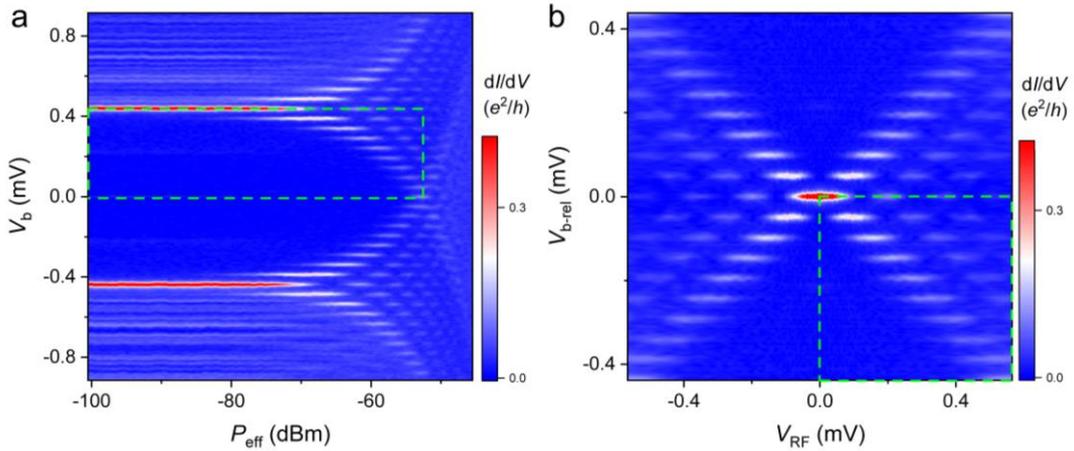

**Figure S6. Selecting the proper data set and converting to the four quadrants. a**, The same data as in Fig. 2c in the main text, but with the effective microwave power $P_{eff} = P - 45.5$ (dBm). The green dashed rectangle displays the data set chosen to perform the 2DFT. The coordinates of this data set were converted to the microwave amplitude $V_{RF}$ and relative bias voltage $V_{b\text{-}rel}$ by setting the $n = 0$ interference fringes to $V_{b\text{-}rel} = 0$ mV, as indicated in **b** by the green dashed rectangle. The data set was further symmetrized to the four quadrants to carry out the 2DFT.

As shown by the green dashed rectangle in Fig. S6a, a clear part of the interference fringes was selected to carry out the 2DFT. The coordinates were converted and further symmetrized to the four quadrants, as displayed in Fig. S6b.

**Step 3**: perform the 2DFT.

To perform the 2DFT, according to the method explained in Section 3, the coordinates of Fig. S6b were further scaled by $me\,V_{b-rel}$ and $meV_{RF}$, and the data were scaled to a maximum of 1. And then the 2DFT can be calculated as a function of $k_\varepsilon$ and $k_A$, as shown in Fig. 2d in the main text. Note that an interpolation of the data was applied when a uniform coordinate spacing was needed.

The analysis of the other data sets of the interference fringes was performed following the same procedure as described above.

**Section 6. Effect of the magnetic field on the LZSM interference.**

In this section, we present the effect of magnetic field on the LZSM interference and the coherence time. Figure S7 shows a comparison of the interference fringes and the corresponding 2DFT patterns for the magnetic fields $B = 0$ T (Figs. S7a and S7b) and 0.2 T (Figs. S7c and S7d; along the nanowire). We can see that the interference fringes became blurred at $B = 0.2$ T, and the observable ovals decreased. Accordingly, the calculated coherence time $T_2$ dropped from ~0.057 ns to ~0.022 ns. We attribute such behavior to the softening of the induced superconducting gap in the InAsSb nanowire when a magnetic field is applied, which induces energy broadening and thus decoherence.

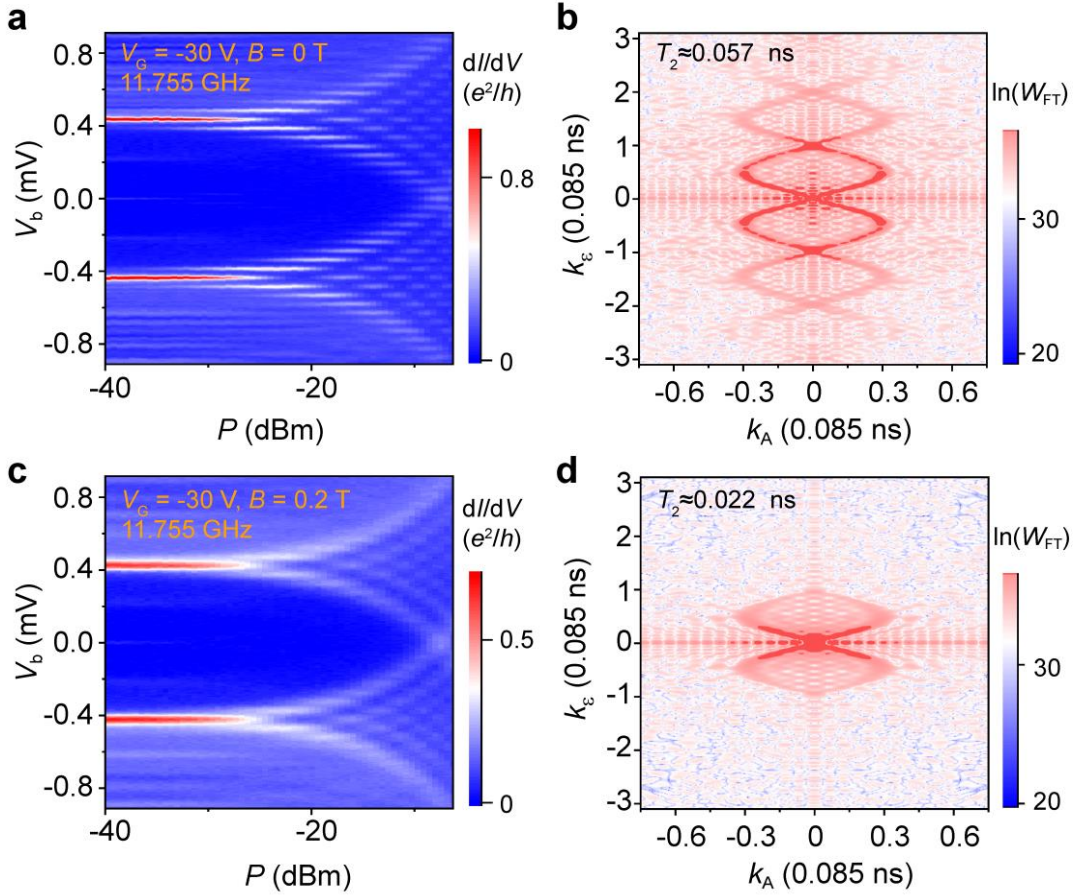

**Figure S7. Effect of the magnetic field on the LZSM interference. a**, Interference fringes at $V_G = -30$ V, $B = 0$ T, and $f = 11.755$ GHz. **b**, 2DFT of **a**. **c**, The same as **a**, but at $B = 0.2$ T. **d**, 2DFT of **c**.

## Section 7. Correlated conductance quantization and critical supercurrent for the second Josephson junction (JJ2).

Figure S8 shows the results for JJ2 (see Fig. 1 of the main text). The conductance $dI/dV$ shown in the upper panel was measured in the normal state at a high-bias voltage: $V_b = 3$ mV (black) and $V_b = 4.5$ mV (red). The quantized conductance plateau at $2e^2/h$ demonstrates the single ballistic channel in the JJ. However, the plateau for $4e^2/h$ can be barely recognized, and the conductance evolves to the plateau of $6e^2/h$ when $V_G$ increases. Such transition from a single channel to three channels could be attributed to the rotation symmetry of the nanowire which induces nearly degenerate 2nd and 3rd sub-bands, as has been observed in InSb nanowires[3]. The bottom panel shows the correlated critical supercurrent $I_C$, where a plateau of ~15 nA was achieved for a single ballistic channel, similar to JJ1 shown in the main text.

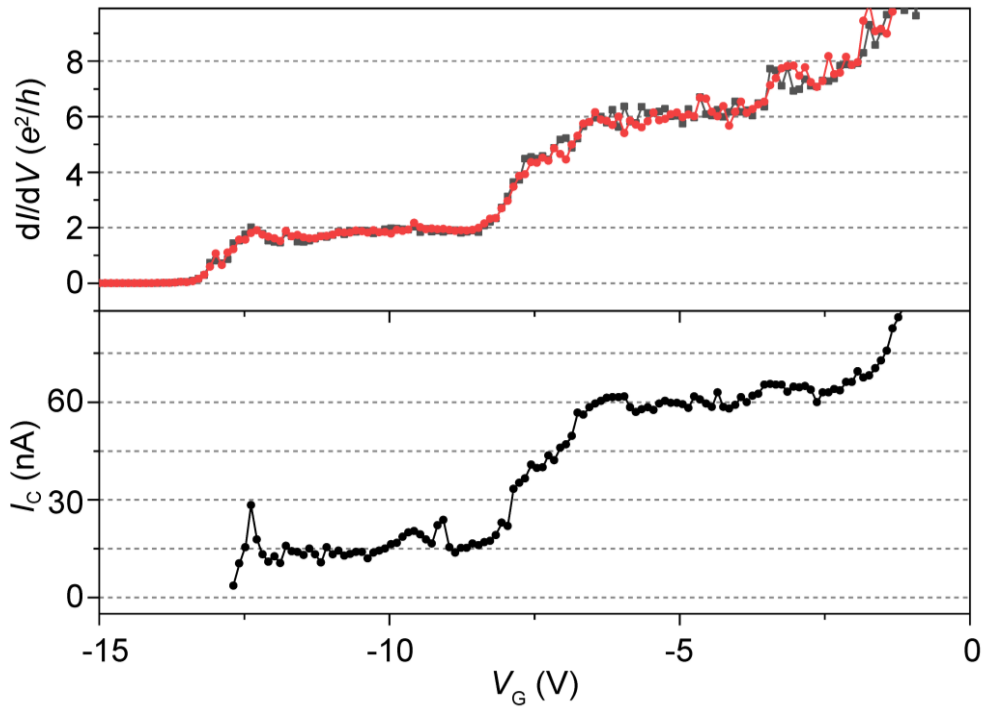

**Figure S8. Correlated conductance quantization and critical supercurrent for JJ2.**

**Section 8. LZSM interference in another typical device.**

LZSM interference can be regularly observed in naturally formed JJs. Figure S9 shows the results of another typical device. An Al gap of ~50 nm was formed during the epitaxial growth of the Al shell due to the shadowing of the dense nanowires, as indicated by the red arrow in the scanning electron microscope image of the device shown in Fig. S9a. The side-gate voltage was set to 0 and a back-gate voltage $V_G$ was applied to control the coupling strength. Figure S9b displays the interference fringes of the Cooper pairs, 2-charges of the 1st-order multiple Andreev reflections, and single charges, similar to the one shown in the main text.

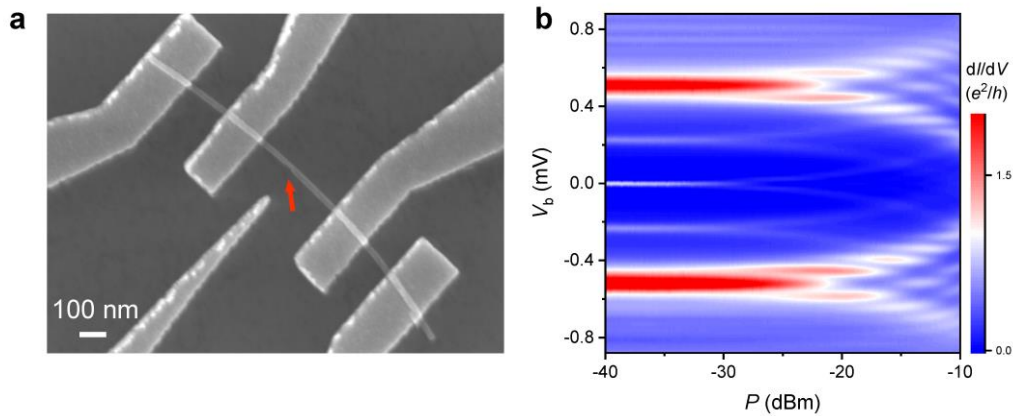

**Figure S9. LZSM interference in another device. a**, Scanning electron microscope image of the device. The red arrow indicates the naturally formed Al gap which functions as the JJ. **b**, Typical interference fringes at a back-gate voltage $V_G = -3.6$ V (300 nm $SiO_2$), a zero side-gate voltage, and $f = 11.3$ GHz.

**Supplementary references:**